\documentclass[prl,aps,amsmath,amssymb,groupedaddress,twocolumn]{revtex4}

\usepackage{graphicx}

\newcommand{\nc}{\newcommand}
\nc{\beq}{\begin{equation}}
\nc{\eeq}{\end{equation}}
\nc{\beqa}{\begin{eqnarray}}
\nc{\eeqa}{\end{eqnarray}}

\def\gsim{\mathrel{\rlap{\lower4pt\hbox{\hskip1pt$\sim$}}
    \raise1pt\hbox{$>$}}}       

\begin{document}

\title{Factorization of unitarity and black hole firewalls}

\author{Stephen~D.~H.~Hsu} \email{hsu@msu.edu}
\affiliation{Department of Physics and Astronomy \\ Michigan State University  }

\begin{abstract}
Unitary black hole evaporation necessarily involves a late-time superposition of decoherent states, including states describing distinct spacetimes (e.g., different center of mass trajectories of the black hole). Typical analyses of the black hole information problem, including the argument for the existence of firewalls, assume approximate unitarity (``factorization of unitarity'') on {\it each} of the decoherent spacetimes. This factorization assumption is non-trivial, and indeed may be incorrect. We describe an ansatz for the radiation state that violates factorization and which allows unitarity and the equivalence principle to coexist (no firewall). Unitarity {\it without} factorization provides a natural realization of the idea of black hole complementarity.
\end{abstract}


\maketitle

\date{today}

\bigskip

\noindent {\bf \large Macroscopic Superpositions}

\bigskip

In this note we elaborate on the role of macroscopic superposition states in black hole evaporation, along the lines of our earlier paper \cite{Hsu} (for related work, see \cite{Nomura,Nomura:2012ex}). The primary object under consideration is the wave function $\Psi$ describing the formation and evaporation of a black hole, including fluctuations in geometry (gravitons). 

Recoil impulses on a black hole from Hawking radiation accumulate into a macroscopic uncertainty in its position of order $\sim M^2$ at late times \cite{Page:1979tc}. Consequently, $\Psi$ after the Page time consists of a superposition of decoherent states describing distinct geometries labeled by $d$: 
\begin{equation}
\Psi = \sum_d \psi_d~~.
\end{equation}
Each $d$ corresponds to a different center of mass trajectory of the black hole, which is related by momentum conservation to the pattern of radiation emission. Because the eventual asymptotic pattern of black hole radiation exhibits statistical fluctuations (i.e., with some spacetime regions containing more energy than others), conservation of momentum requires distinct spacetime geometries in $\Psi$ at late times. 

Typical analyses (e.g., the standard nice-slice construction, or the firewall argument of AMPS \cite{Almheiri:2012rt,Almheiri:2013hfa,Brau,Mat}) of the black hole information problem are conducted on a fixed spacetime background. Approximate unitarity of the evaporation process on an individual $\psi_d$ requires
\begin{equation}
\label{U}
\psi_d (f) \approx U \psi_d (i)~~,
\end{equation}
where $i,f$ denote early and late times, and the symbol $``\, \approx \,"$ means that the final (radiation) state $\psi_d (f)$ is almost pure, or has small von Neumann entropy relative to the black hole.  We refer to this condition as {\it factorization of unitarity} over the decoherent spacetimes. Factorization is a much stronger condition than $\Psi (f) = U \, \Psi (i)$, which only requires unitarity when {\it all} branches are taken into account. It does not appear that AdS/CFT duality requires factorization: the full development of $\Psi$ is presumably represented in the CFT, and only overall unitarity can be inferred.

An objection to the importance of macroscopic superpositions to the information problem is that there is less information in the coarse grained position or even trajectory (sequence of positions) of the black hole than in the radiation. From this perspective one should be able to neglect the superposition of spacetimes and demand approximate unitarity branch by branch -- in other words, impose factorization. Below, we show that the firewall argument depends sensitively on factorization. 
Once macroscopic superpositions are taken into account, the required deviation of near-horizon modes from the inertial vacuum state becomes extremely small.

\bigskip

\noindent {\bf \large Factorization and Firewalls}

\bigskip

Consider the AMPS argument for firewalls \cite{Almheiri:2012rt,Almheiri:2013hfa}. Let $A$ = modes in black hole interior, $B$ = near horizon exterior modes, $C$ = early radiation. The subscript $d$ (e.g., $B_d$) denotes those modes on the specific spacetime background $d$. Let the symbol $``~ \& ~"$ mean {\it strongly entangled with}.

\smallskip
1. equivalence principle implies  $A_d \, \& \, B_d$  

2. factorization of unitarity implies $B_d \, \& \, C_d$.
\smallskip

Because $B_d$ cannot be strongly entangled with both $A_d$ and $C_d$, we must give up either the equivalence principle or the factorization of unitarity (i.e., approximate unitarity on $d$ -- as we noted above, this is {\it not} the same as overall unitarity of $\Psi$). AMPS choose to give up the equivalence principle in favor of a firewall (deviation from vacuum state as seen by inertial observer near the horizon). But, one could instead give up factorization of unitarity in favor of the weaker assumption that unitarity holds only after all branches are taken into account.

Strictly speaking, the purification of the global radiation state by late quanta only requires $B \, \& \, C$, not $B_d \, \& \, C_d$. Here global states $B$ or $C$ mean summing over all geometries \cite{FN2}. As we discuss below, the entanglement between $B_d$ and $C_d$ could be quite small, requiring only a negligible deviation from $A_d \, \& \, B_d$. Indeed, the late time uncertainty in black hole position is $\Delta x \sim M^2$, which means there are (up to a coefficient determined by experimental sensitivity) at least $N_d \sim M^6$ distinct branches $d$, counting just by CM position at one instant in time, not full trajectory over the entire evaporation. Counting trajectories (i.e., sequences of positions) leads to $N_d \sim \exp  M^2 $. In the absence of degenerate particles of different species, essentially all of the information in an individual particle emission is captured by the recoil of the black hole (i.e., kinematics identify the mass, energy, momentum and spin of the particle), so the number of possible coarse grained black hole recoil trajectories is similar to the number of possible radiation states -- the two could be equal, up to a coarse graining factor. If the entanglement between $B$ and $C$ is spread over a large number of $B_d$ and $C_d$, the resulting entanglement between any pair $B_d$ and $C_{d'}$ could be extremely small. 

We can represent the global state as follows, neglecting for the moment the near-zone modes $B$. The subscript denotes subsets of the Hilbert space describing individual decoherent geometries:
\begin{equation}
\Psi (t) ~\sim~ ( C_1, C_2, \cdots, C_{N_d} ; A_1, A_2, \cdots A_{N_d}) ~~.
\end{equation}
The dimensionality of the $A$ Hilbert space is ${\rm dim} \, H(A) \sim N_d (t) \exp M(t)^2$. As the holes evaporate $M(t)$ decreases but $N_d (t)$ increases -- there are more and more holes in different locations, each with $\sim  \exp M(t)^2$ possible future radiation patterns. For $N_d (t) \sim \exp (M^2 - M(t)^2)$, we have ${\rm dim} \, H(A) \sim \exp M^2$, independent of time. This may seem counterintuitive; the reason ${\rm dim} \, H(A)$ does not decrease is that as the evaporation proceeds the number of patterns of emitted radiation increases, and so does the number $N_d (t)$ of possible realized trajectories of the hole. In fact, ${\rm dim} \, H(C) \sim \exp (M^2 - M(t)^2) \sim N_d (t)$, up to an overall coarse graining factor. See \cite{FNQ} for additional discussion in the context of a simple qubit model.

Non-factorization means that the radiation emitted at time $t$ depends on the {\it entire} global state $A(t)$ -- that is, the internal state $A_d$ can influence radiation emitted on branch $d' \neq d$. This violates locality in a certain sense, linking states on different geometries (it is quite different from a burning lump of coal: $A_d$, the internal state of the lump $d$, cannot significantly influence radiation emitted by the lump $d'$ -- i.e., the lump at a different position). The important consequence is that under the non-factorization assumption, density matrices describing $B$, $C$ or $B_d$, $C_d$ require tracing over all of $A$: e.g., $\rho_{BC} = {\rm tr}_A  \vert \Psi \rangle \langle \Psi \vert$.

Now consider the subadditivity inequality 
\begin{equation}
\label{bound}
S_{A_d} + S_{C_d} \leq S_{A_d B_d} + S_{B_d C_d}~. 
\end{equation}
If the additional radiation $B_d$ serves to purify $C_d$, then $S_{B_d C_d} < S_{C_d}$, which leads to a contradiction since $B_d$ purifies $A_d$, so that $S_{A_d B_d} < S_{A_d}$. However, we show below that both $B_d C_d$ and $C_d$ are nearly maximally mixed. Therefore, taking $A_d$ and $B_d$ to be entangled pairs of in- and out-going Hawking quanta, we have $S_{B_d C_d} - S_{C_d} \approx S_{B_d} = S_{A_d}$, up to an exponentially small correction. As a result, the near horizon region $A_d B_d$ need only deviate from the inertial vacuum by an exponentially small perturbation.

To obtain the necessary result, we use theorem III.3 of \cite{HLW}: let $X = B_d C_d$ or $C_d$, which are very small subsets of $ABC$. The states in $A$ are behind the horizon and must be traced over to obtain a description of $X$. The dimensionality of $A$ is much larger than that of $B_d C_d$ or $C_d$. Then, Theorem III.3 of \cite{HLW} implies that, with overwhelmingly high probability, a random pure state $\phi \in ABC$ leads to a density matrix $\rho_X$ with nearly maximal entropy. The likely deviation from maximum entropy will be exponentially small.

We summarize this result as follows. The firewall argument requires that we work on a particular geometry $d$ since we wish to invoke the equivalence principle near the (i.e., a specific) horizon. Even if the observer outside the hole can make perfect measurements on every degree of freedom outside the horizon, he is still ignorant of the state of the interior $A$. Tracing over $A$ yields (due to entanglement) a mixed state description for $C_d$ and $B_d C_d$. For a typical pure state describing the full system $ABC$ (i.e., not imposing factorization, which corresponds to a subset of negligible measure), and given the fact that $A$ is of much higher dimensionality than the radiation Hilbert space consistent with a particular $d$ geometry, one can show that $C_d$ and $B_d C_d$ have nearly maximal entropy. Consequently, the purification of the early radiation obtained by the emission of $B_d$ is negligible, and the required deviation of $A_d B_d$ from a pure state that can be deduced from (\ref{bound}) is exponentially small. 

We can also formulate this discussion in terms of single modes. In the notation of \cite{Almheiri:2013hfa}, let $b$ be a late-time Hawking mode, $\tilde{b}$ its interior partner, and $e_b$ an early time Hawking mode entangled with $b$. Roughly speaking, the unitarity assumption implies that there is some early time mode $e_b$ with the property that $( b \, e_b )$ form a pure state. But this conflicts with the (no drama, or equivalence principle) assumption that $(b \, \tilde{b})$ form a pure state. The resolution is that $( b \, \tilde{b} )$ are excitations relative to the vacuum of a particular $d$ spacetime, whereas unitarity only holds across all $d$ branches -- there is no $( b \, e_b )$ pair residing solely on a particular $d$ branch. If such $( b \, e_b )$ pairs existed, we would have found that $B_d C_d$ becomes increasingly pure as the hole evaporates. 

For a typical pure state $\phi \in ABC$, the entanglement between $A$ and $C_d$ or $B_d C_d$ is nearly maximal \cite{HLW}. This entanglement is transferred to the radiation $C_{d'}$ on {\it other} geometries $d'$. Ultimately, if the black hole is in a typical state during intermediate times, the final radiation state $C$ will exhibit strong entanglement across all $d$ sectors, motivating the ansatz we discuss next.

\bigskip

\noindent {\bf \large No Firewall Ansatz}

\bigskip

Consider the following radiation density matrix in the $\exp S$ dimensional Hilbert space \cite{Papadodimas:2012aq}:
\begin{equation} \label{rho}
\rho = \rho_{\rm Hawking} + \rho_{\rm correction}~,
\end{equation}
where 
\begin{equation}
\rho_{\rm Hawking} ~=~ {\rm diag} ( 1/N, \cdots, 1/N ) ~ \sim~  e^{-S} ~\bf{1}~,
\end{equation}
and
\begin{equation}
\label{correction}
\rho_{\rm correction} ~\sim~ e^{-S} ~\bf{Q}~. 
\end{equation}
$\bf{Q}$ is a matrix with typical entries ${\cal O} (1)$. The form of $\rho_{\rm correction}$ reflects its origin in small quantum corrections to Hawking's semiclassical result. For appropriate $\bf{Q}$, it is possible that ${\rm Tr} \, \rho^2 = 1$ (i.e., $\rho$ is a pure state), despite the maximally mixed nature of $\rho_{\rm Hawking}$. Note that the specific geometry $d$ is {\it determined} by the radiation record -- i.e., the pattern of recoil kicks. So the decohered $d$ sectors are subsets of the $\exp  S$ dimensional Hilbert space. 

This ansatz can be constructed to exhibit strong correlations {\it between} subspaces with different values of $d$. If most entries in $\bf{Q}$ are ${\cal O} (1)$, the entanglement of $B_d$ can be {\it spread over all of} $C$. If this is the case, then only a tiny fraction of the entanglement is between $B_d$ and $C_d$. This small amount of entanglement does not conflict with $A_d \, \& \, B_d$ -- it requires only a small deviation from a pure state $A_d B_d$ (the inertial vacuum).  

The physical picture is that quantum gravitational effects cause distinct $d$ geometries to become entangled. {\it This sounds less exotic if we realize that different $d$ geometries merely correspond to macroscopically different radiation patterns.} Quantum entanglement across different radiation patterns is often proposed as a way to encode black hole information. Some possible quantum mechanisms were discussed in \cite{Hsu}. 

Note that if the radiation emitted on geometry $d$ is dependent on the global internal state $A$ of {\it all} black holes (i.e., across all $d'$ geometries), there is no natural sense of {\it locality} -- $A_d$ can influence $B_{d'}$ or $C_{d'}$, even at some location beyond the horizon at which the Hawking pair $(b,\tilde{b})_{d'}$ are created. The notion of spacelike separation is only defined with reference to a particular geometry $d$ and does not generalize to interactions between $d$ and $d'$. In fact, the spread $\Delta x \sim M^2$ in locations of the hole extends far beyond the near horizon region of any specific hole. This kind of non-locality seems to be required by Mathur's Theorem \cite{Mat} -- if the horizon region generates the usual Hawking pairs, then influences at larger distances are required to solve the information problem. It also has some similarities to the ``wormhole'' proposal of Maldacena and Susskind \cite{Maldacena:2013xja}.

Of course, we don't know whether the ansatz $\rho$ in (\ref{rho}) is correct -- perhaps factorization in fact holds, despite the fact that this would require highly atypical states $ABC$. Nevertheless, the ansatz gives us a specific example that evades the AMPS firewall argument. One could regard the firewall paradox as a hint that correlations across decoherent branches are crucial for unitarity -- in particular, for maintaining both unitarity and the equivalence principle.

\bigskip

\noindent {\bf \large Complementarity}

\bigskip

Black hole complementarity attempts to reconcile the seemingly contradictory experiences of Alice, who falls behind the horizon, and the outside observer who can in principle verify unitarity. With many branches, it is natural to have different observers with contradictory subjective experiences. For example, in the ansatz considered above, the global evolution of the system is unitary, even though many Alices will experience falling through the horizon. {\it Could this be the basis for black hole complementarity?} The usual nice slice + no cloning argument is evaded because the nice slice is specific to a $d$ sector, whereas the ``cloned state'' of Alice (in the global radiation $C$) is spread over many $d$ sectors.

Recall the results of Popescu et al. \cite{Winter} using Levy's lemma (concentration of measure in high dimensions). Consider a pure state $\Psi$ subject to a linear energy constraint. Let the Hilbert space consist of two disjoint subspaces $X$ and $Y$, $d_Y \gg d_X$. Then $\rho_X = {\rm Tr}_Y \Psi$ is approximately thermal. The experimental sensitivity required to determine that $\rho_X$ is not exactly thermal is $\sim 1/ d_Y^{1/2} \sim \exp - S$. Roughly (norm = trace norm):
\begin{equation}
\label{P}
{\rm Prob} \left( || \rho_X - \rho_{\rm therm} || > \epsilon    \right) ~ <~  \exp ( - \epsilon^2 d_Y )~.
\end{equation}
Thus an observer in sector $d$ will {\it not} observe the radiation entropy decrease to nearly zero (i.e., will not insist that $B_d \, \& \, C_d$), {\it unless that observer has sufficient experimental power that she would be sensitive to other d sectors}. That is, with $\exp - S$ sensitivity, one can overcome the decoherence separating different $d$ sectors, which have wave function overlap $\sim \exp - S$. Without this level of precision, the observer will conclude that the radiation is thermal (mixed) and not insist that $B_d \, \& \, C_d$ at late times. 

To verify that a state is pure one really has to measure it ``all at once''; measuring only part risks perceiving it as a mixed state. But in the case of black hole evaporation, to measure ``all'' of it would require the ability to measure across $d$ sectors. The version of complementarity proposed here is that an Alice who experiences falling through a (particular) horizon is by definition not sensitive to other $d$ branches. She also, therefore, cannot determine whether the radiation is in a pure state, or whether $B_d \, \& \, C_d$. An observer that can determine purity of the radiation knows about other $d$ branches and cannot experience falling through a particular horizon. Note that we have stressed the necessity of exponentially fine experimental sensitivity to detect decoherent branches, but an alternative is carefully tuned measurement operators involving macroscopic superpositions \cite{HsuBH}.

\bigskip

\noindent {\bf \large Conclusions}

\bigskip

The quantum evolution of a complex pure state typically leads to a superposition of decoherent semiclassical states. In the case of black hole evaporation one obtains a superposition of spacetime geometries because the Hawking radiation inevitably exhibits fluctuations in energy and momentum density over different regions. Firewall and information paradoxes result from the non-trivial assumption of factorization: approximate unitarity on each decoherent geometry. Global unitarity is a much weaker condition than factorization. Quantum correlations between geometries can plausibly resolve the information paradoxes, although specific dynamical mechanisms are still not understood.

\bigskip

\noindent {\bf \large Acknowledgements}

\bigskip

The author thanks Don Marolf, Samir Mathur, Joe Polchinski, John Preskill and Chiu Man Ho for discussions. This work was supported by the Office of the Vice-President for Research and Graduate Studies at Michigan State University.


\bigskip



\begin{thebibliography}{99}
\bibitem{FNQ} Consider the evolution of the ``black hole'' interior state $A$ as it emits Hawking radiation: $\vert A_0 \rangle \rightarrow \vert A'_0 \, (- +) \rangle + \vert A''_0 \, (+ -) \rangle \equiv \vert A_+ , + \rangle + \vert A_- , - \rangle$, where $(- +)$ represents the Hawking state with an infalling $-$ qubit and an outgoing $+$ qubit, and $(+ -)$ the opposite. $A_+$ is the interior state after a $+$ mode is emitted, $A_-$ corresponds to a $-$ mode emitted. Clearly ${\rm dim} \, H(A_\pm) \equiv {\rm dim} \, H(A_1)= 2 \,  {\rm dim} \, H(A_0)$, and after $n$ quits are emitted: ${\rm dim} \, H(A_n) = 2^n \, {\rm dim} \, H(A_0)$. In other words, the infalling qubits increase the entropy of $A$ (enlarge the set of possible states $A$ could be in) by an amount equal to the entropy of the outgoing radiation. If we impose that the original entropy of the hole $A_0$ is reduced by an amount equal to that of the outgoing radiation, the total entropy of the interior $A$ (and therefore the dimensionality of its Hilbert space) is constant during the evaporation.
\end{thebibliography}
\end{document}